\documentstyle[aps,prl,epsfig,twocolumn]{revtex}
\begin{document}

\wideabs{
\title{Structural instability of vortices in Bose-Einstein condensates}

\author{Juan J. Garc\'{\i}a-Ripoll$^{\dag}$, Gabriel Molina-Terriza$^{\ddag}$,
  V\'{\i}ctor M. P\'erez-Garc\'{\i}a$^{\dag}$, Lluis Torner$^{\ddag}$}

\address{$^{\dag}$Departamento de Matem\'aticas, E. T. S. I. Industriales,
  Universidad de Castilla-La Mancha, 13071 Ciudad Real, Spain. \\
  $^{\ddag}$ Laboratory of Photonics, Department of Signal Theory and
  Communications, \\
  Universitat Polit\'ecnica de Catalunya, Jordi Girona 1, UPC-D3, \\
  Barcelona, ES 08034, Spain.}

\date{\today}

\maketitle

\draft

\begin{abstract}
  In this paper we study a gaseous Bose--Einstein condensate (BEC) and show
  that: (i) A minimum value of the interaction is needed for the existence of
  stable persistent currents. (ii) Vorticity is not a fundamental invariant of
  the system, as there exists a conservative mechanism which can destroy a
  vortex and change its sign. (iii) This mechanism is suppressed by
  strong interactions. 
\end{abstract}

\pacs{PACS number(s): 03.75.Fi, 05.30.Jp, 67.57.De, 67.57.Fg}
}

Superfluidity is an intriging concept which dates back to the discovery of
He--II \cite{Kapitza}, and which is by now textbook material. Among the
properties of a superfluid there are irrotational flows, quantization of
circulation and stability of persistent currents. The theoretical disadvantage
of He--II is that the strong interactions which take place in the liquid
prevent a simple microscopical description of the system, and make it
impossible to clearly untangle the ultimate cause of superfluidity.

The achievement of Bose-Einstein condensation with ultracold atomic gases has
revived the interest on a theoretical description of superfluidity. The first
reason is that Bose--Einstein condensation has long been thought to be the
cause for the superfluid behavior in He--II \cite{Feynmann}. But the second
reason is that gaseous condensates can be accurately described with simple
mean field models that can be subject of both analytic and numerical
study \cite{Dalfovo}.

In the last few years, the superfluid nature of these condensates has got
experimental demonstrations by several means. One of the most striking features
of superfluids is the quantization of circulation, which leads to the existence and
 {\em stability} of vortices even with the lack of external 
 forcing. Vortices have been
theoretically predicted to be present in gaseous condensates with repulsive
interactions \cite{Fetter-review,feder,butts,rokhsar,vortices}, a prediction
which has been confirmed in recent experiments \cite{vortex-exper,ENS,Keterlee} where
vortices have been produced and shown to be, at least, long--lived.

The stability of vortices as seen in experiments and in numerical studies, has
lead us to consider vortices and their topological charge as robust entities.
In this work we prove the surprising fact that a weakly interacting condensate
in a stationary trap cannot host stable persistent currents of well defined
vorticity, showing that nonlinearity is therefore essential to sustain vortices. 
 We show what is the mechanism for the structural destabilization of vortices and propose a
way to verify this prediction with in current experiments.

{\bf The model.-} In this work we study a dilute gaseous BEC in the
zero--temperature limit. This system is ruled by the Gross--Pitaevskii mean
field equation \cite{Dalfovo}
\begin{equation}
  \label{eq:gpe}
  i\hbar \frac{\partial \Psi}{\partial \tau} =
  \left[-\frac{\hbar^2}{2m}\triangle + V({\bf r}) + \frac{4\pi a_S\hbar}{m}|\Psi|^2\right]\Psi,
\end{equation}
where $\triangle = \sum_{j=1}^n \partial^2/\partial r_j^2$ is the
$n$-dimensional Laplacian, $V({\bf r}) =
\frac{1}{2}m\omega^2\sum_{j=1}^n\lambda_jr_j^2$ is the external potential which
confines the condensate and $a_S$ is $s$--wave scattering length for the binary
collisions within the condensate. We will focus on condensates with $a_S \geq
0$, which represents the repulsive interaction between bosons. The norm of the
wavefunction corresponds to the number of bosons in the trap, $N =
\Vert\psi\Vert^2$.

Due to the quantum nature of the system we are studying, the velocity of
the fluid can be derived from the macroscopic wavefunction of the condensate
\begin{equation}
  \label{eq:velocity-field}
  {\bf v} = \frac{i\hbar}{2m}(\Psi\nabla\bar{\Psi} - \bar{\Psi}\nabla\Psi)({\bf r})
  = \frac{i\hbar}{2m}\rho({\bf r})\nabla\theta({\bf r}).
\end{equation}
Here we have separated the wave function into its modulus, $|\Psi| =
\sqrt{\rho}$, and its phase, $\theta =\arg\Psi$.

From now on we will use adimensional quantities to ease the analysis. To do so
we define a new set of units based on the trap characteristic length,
$a_0=\sqrt{\hbar/m\omega}$, and period, $T=1/\omega$ defined as $x_j = r_j/a_0,
t = \tau/T$ and a new wavefunction $\psi({\bf r},t) \equiv \Psi({\bf x},\tau)$ satisfying
\begin{equation}
  \label{eq:gpe2}
  i \frac{\partial \psi}{\partial t} =
  \left[-\frac{1}{2}\triangle + \frac{1}{2}\sum_j\lambda_j^2x_j^2 + U|\psi|^2\right]\psi,
\end{equation}
where $U = 4\pi a_{S}/a_0$. The value of $NU$
measures the relative importance of the nonlinear term $U|\psi|^2$.  Through
this paper we will call a \emph{weakly interacting condensate} a system in
which $NU \leq 1$ as oposed to a \emph{strongly interacting condensate} in
which $NU \gg 1$ \cite{note}.

The energy of the system is given by
\begin{equation}
{\cal{E}}[\psi ] = \int \bar{\psi}\left[ -\frac{1}{2}\triangle +V\left( \mathbf{x}\right)
+\frac{U}{2}\left| \psi \right| ^{2}\right] \psi d{\bf x}, \label{energy}
\end{equation}
which is a conserved quantity.

In this work we will consider a BEC placed in a tighly confining pancake type trap \cite{MIT-new},
 for which the profile of the solution along the confined dimension is very well approximated by a 
 gaussian function \cite{Perez-Garcia98}.
The procedure for reducing the Gross-Pitaevskii equation to an effective two--dimensional equation 
is described in detail in Ref. \cite{us-rota}. The only relevant difference with Eq. (\ref{eq:gpe2}) reduced to 
two spatial dimensions is that the nonlinear term is reduced by a factor related to the trap asymmetry. 
We  will consider then that our effectively two--dimensional system is described by Eq. (\ref{eq:gpe2})
 but with a new $U = \alpha 4\pi a_S \alpha/a_0$ \cite{note2}.

{\bf Noninteracting systems.-} A persistent current is a stationary solution in
which the velocity field is nonzero, ${\bf v} \neq 0$, and remains still,
$\partial_t{\bf v}=0$, even without external forcing. Therefore, any solution
of Eq. (\ref{eq:gpe}) corresponding to a persistent current must have a
position dependent phase. To ease the analysis we will write such stationary
solution as $\psi_{\mu}({\bf x},t) = \phi_{\mu}({\bf x}) e^{i\mu t} =
\psi_{\mu,r} + i\psi_{\mu,i}$, where $\psi_{\mu,r}$ and $\psi_{\mu,i}$ are real
functions.

We will first consider the noninteracting case ($U = 0$), in which the
persistent current satisfies
\begin{equation}
  \label{eq:linear}
  H \phi_{\mu} = \mu \phi_{\mu},
\end{equation}
with the Hamiltonian $H=-\frac{1}{2}{\triangle} + V({\bf x}).$ Remarkably, Eq.
(\ref{eq:linear}) implies that $\psi_{\mu,r}$ and $\psi_{\mu,i}$ must be
eigenstates of $H$ with the same eigenvalue. Therefore, {\em in noninteracting
systems, persistent stationary currents can exist only if the spectrum of the
Hamiltonian is degenerate}.

Furthermore, degeneracy is a very specific situation which can be broken by
almost any perturbation --e.g. any small asymmetry of the confining
potential--, which leads us to conclude that persistent currents in
noninteracting systems are {\em structurally unstable}, a term which will be
precised later.

To understand what happens to the quantum fluid in the noninteracting regime,
let us consider a simple, two dimensional harmonic oscillator whose three lowest
energy eigenstates of $H$ are $\phi_0 \propto \exp\left(-\frac{x^2}{2a^2} - \frac{y^2}{2b^2}\right),$
$  \phi_{x} \propto x \phi_0$, $\phi_y  \propto  y\phi_0$, where we make use of the adimensional natural 
widths of the oscillator, $a=1/\sqrt{\lambda_x}$, $b=1/\sqrt{\lambda_y}$.

In the case without interaction ($U=0$) and radial symmetry ($a=b$), vortices
are solutions of the type $\psi(x,y,t) = \sqrt{N/2} \left[\phi_x + i \phi_y\right] e^{i\mu t},$
which is an eigenfunction of $H$. The analogous of this current for the asymmetric
case ($a \neq b$) is $\psi(x,y,t) \left(x\sqrt{N_x} + iy \sqrt{N_y} e^{i(\mu_y-\mu_x) t} \right)e^{i\mu_at} 
 \phi_0$. Due to the existence of a time dependent phase between both components, the
solution is not stationary and evolves far from the initial configuration. In
fact, for $t = \pi/(\mu_y-\mu_x)$ the relative phase leads to a complete
inversion of the topological charge of the vortex from $m=+1$ to $m=-1$ passing through an intermediate state
where a dark line breaks the condensate into two pieces. 

{\bf Structural stability.-} Thus, for the noninteracting case: (i) There are
no stationary vortex solutions in any stationary asymmetric trap (all
eigenfunctions are real). (ii) Vortex type solutions are unstable in those
traps since they follow an evolution which leads to an inversion of the
topological charge.

The vortex reversal can be induced by any small perturbation
that breaks the degeneracy, such as small asymmetries or even spatial noise. This is an
example of {\em structural instability}.  The important difference between {\em
stability} referred to perturbations of the initial conditions and {\em
robustness} referred to perturbations of the governing evolution equations,
is not always properly appreciated, but it is a key ingredient of the
phenomenon discovered here. 

It is important to stress that our previous finding implies that 
there exists a fundamental limitation on the stability of
vortices, which makes that as soon as the forcing is removed, these currents
will persist only for a period of time that depends on the strength of the
interactions and the amount of perturbation.

{\bf Role of interactions.-} Let us now consider the interacting case and
look for the simplest stationary currents on a two--dimensional condensate in
the very weakly interacting limit. In this limit the ground state of
the condensate is almost equal to the ground state
of the linear problem $\psi_0$, and for the vortices we
may use the simple ansatz $\psi_v = \sqrt{N_x} \phi_x + i\sqrt{N_y} \phi_y$,
where $N_x$ and $N_y$ are variational parameters and can be regarded as the
number of atoms on each dipole state. The energy of our ansatz is
obtained substituting into Eq. (\ref{energy})
\begin{eqnarray}
  \label{eq:energy}
  E &=& \mu_xN_x + \mu_yN_y + \frac{U}{8\pi a b} \left(3N_x^2 + 3N_y^2 + 2N_xN_y\right),
\end{eqnarray}
where $\mu_{x}=\frac{3}{2}\lambda_{x}+\frac{1}{2}\lambda_{y}$
and $\mu_{y}=\frac{3}{2}\lambda_{y}+\frac{1}{2}\lambda_{x}$
are the
eigenvalues of the $\phi_{x,y}$ basis functions.

\begin{figure}
  \begin{center}
  \epsfig{file=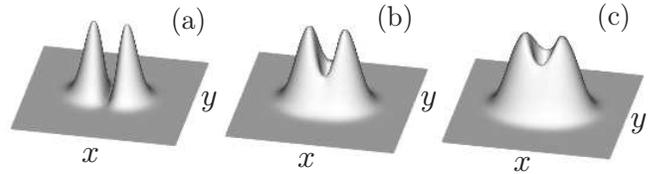,width=\linewidth}
  \end{center}
  \caption{\label{fig:states}
    (a) One of the dipole solutions in a very weakly interacting
    condensate. (b-c) Density plots of stationary vortices for $UN\simeq 14, 31$
    respectively. All plots expand the spatial region
    $[-3,3]\times[-3,3]$ and correspond to a trap with
     $\lambda_y/\lambda_x = a^2/b^2 = \sqrt{2}$.}
\end{figure}

We have to search the critical points of (\ref{eq:energy}) over the line
$N=N_x+N_y$.  When $U=0$ the energy functional has only two such critical
points which correspond to the $\phi_x$ and $\phi_y$ stationary states. When
$U\neq 0$ we still find those two dipole solutions with energies given by
\begin{eqnarray}
  \label{eq:states}
  E_x &=& \left(\mu_x + \frac{3UN}{8 \pi a b}\right)N,\quad N=N_x,N_y=0\\
  E_y &=& \left(\mu_y + \frac{3UN}{8 \pi a b}\right)N,\quad N=N_y,N_x=0.
\end{eqnarray}

Furthermore, {\em if and only if the interaction exceeds a critical value}
\begin{equation}
  \label{eq:critical-U}
  UN >  2\pi a b |\mu_x - \mu_y| = \sigma,
\end{equation}
{\em there exists a third solution, which is a deformed vortex}, and whose
energy and populations are
\begin{mathletters}
\begin{eqnarray}
  \label{eq:vortex}
  E^{(v)} & = & \left(\frac{\mu_x+\mu_y}{2}+\frac{UN}{4\pi ab}\right)N
  -\frac{\pi a b (\mu_x-\mu_y)^2}{2U},\\
  N_{x,y}^{(v)} & = & \frac{N}{2} - \frac{\pi a b}{U}\left(\mu_{x,y}-\mu_{y,x}\right).
\end{eqnarray}
\end{mathletters}

According to these equations we classify the $(N,U)$ space into several
regions. For $UN \leq
\sigma$, no vortex is found. Above $UN \geq \sigma$ not only the vortex exists but it is also
energetically more favorable than the dipoles therefore preventing the possibility that the cloud
spontaneously splits into two pieces.

{\bf Numerical simulations.-} To study the sensitivity of the system to finite
(but may be ``small") perturbations we have proceeded as follows. First we have
found the stationary vortex profile for given asymmetries of the trap and interaction
values [See e.g. Fig.  \ref{fig:states}(b-c)] using the numerical method
developed in Ref. \cite{sobolev}.  Next we have used this stationary solution
as initial data for a numerical simulation of the dynamics of Eq. (\ref{eq:gpe2})
in a perturbed system where the perturbation used consisted on a change of the trap parameters.

For values of $UN < \sigma$ we were not able to find any vortex solutions, in
agreement with the prediction from the variational analysis.

\begin{figure}
  \begin{center}
  \epsfig{file=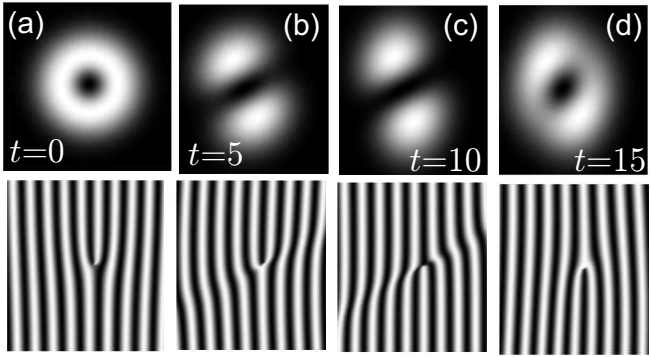,width=\linewidth}
  \end{center}
  \caption{\label{fig:weak}
    Rupture of a symmetric vortex ($\lambda_x = \lambda_y = 1$, $UN \simeq 2.2$)
    placed in a trap with $\lambda_x = 1, \lambda_y =0.77$.
     Shownare density plots and phase (interference) plots on the spatial
     region $[-2.5,2.5]\times[-2.5,2.5]$ (adimensional units).}
\end{figure}

\begin{figure}
 \begin{center}
 \epsfig{file=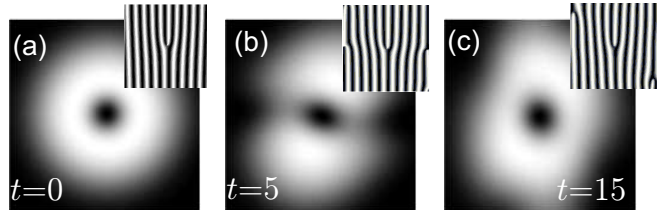,width=\linewidth}
 \end{center}
  \caption{\label{fig:robust}
  Stability of a symmetric vortex ($\lambda_x = \lambda_y = 1$, $NU \simeq 30$)
  placed in a trap with $\lambda_x = 1, \lambda_y = 0.71$ ($\sigma \simeq 2.2$).
  Shownare density plots and phase (interference) plots on the spatial
  region $[-3,3]\times[-3,3]$ (adimensional units).}
\end{figure}

However, we find that for medium or small $UN$, where vortices already exist
and are suposed to be linearly stable, the application of finite perturbations leads
to the destabilization of the vortex and the generation of a dynamics with some
analogues with the vortex inversion mechanism discused above in the framework of the
noninteracting system (see Fig. \ref{fig:weak}). The smaller the value of $UN$,
the smaller the amount of perturbation which is sufficient to destabilize the system.
For moderately large values of the self interaction structural stability adds to the
dynamical stability that was previously known to exist
\cite{Fetter-review,feder,vortices}. In this case vortices are stable and keep
their topological charge even when oscillations are induced to the vortex
profile (See e. g. Fig. \ref{fig:robust}) by strong perturbations of the
confinement. In this region, neither small noise nor changes in the confinement
are able to break the vortex.

We have also verified that the phenomenon of vortex charge flip is not exclusive of vortices in asymmetric traps. In fact,
as the general analysis of the linear problem suggests, any symmetry breaking perturbation should lead to
a simmilar phenomenon. To test this idea we have studied numerically the effect of a thin and intense pinning potential
placed asymmetrically with respect to the trap center and verified that the same mechanisms involving topological 
charge inversion are present. This interesting result adds value to the prediction since the study of the effect of 
pinning potentials on vortices is one of the goals of several experimental groups (see e.g. \cite{Keterlee}). In fact
many other dynamical features are found and will be reported in detail in a future paper \cite{PPP}.

{\bf Discussion and conclusions.-} In this work we have studied the role of
interactions on the properties of vortices of a gaseous, trapped Bose--Einstein
condensate. Our main conclusion is that vortices in a weakly interacting
condensate are structurally unstable and a critical value of the
interaction is required for the condensate to host stable stationary currents.

Although the variational approach states that for any $NU > \sigma$ vortices
exist and become energetically favourable against dipole states, our numerical 
simulations suggest that there exists a finite range of weak interaction values in which
the vortex is close to structural instability and breaks under small perturbations.

It is important to stress that this phenomenon may be experimentally observed using either asymmetric pancake traps \cite{MIT-new}
or Feschbach resonances \cite{Cornell,Feschbach2}.

As we have discussed above, the effective interaction in a quasi two--dimensional system is reduced by a factor depending
 on the asymmetry of the trap \cite{us-rota}. Using the numbers of the pancake trap recently depeloped by MIT group to test low dimensionality 
effects \cite{MIT-new} we get that $U_{2D} \simeq 3\times 10^{-2}$. This estimate implies that a number of particles about 
$N \simeq 10^4$ would lead to a value of $UN = 30$ which is in the parameter range in which the phenomenon should be 
observable. It is interesting that the trap used in \cite{MIT-new} is already asymmetric
since $\omega_x/2\pi =$ 10 Hz and $\omega_y/2\pi =$ 30 Hz, i.e $\sigma \sim 10$. These numbers mean that the vortex 
topological charge inversion should be observable by just placing it into one such trap (using e.g. a rotating pinning 
such as the ones used by the same group and eliminating it once the vortex is generated). We have verified numerically
that the vortex charge flip is present in this parameter range and even for larger values of $N$. The fact that in the experimental setup
the trasverse asymmetry is not ``small" helps in making the phenomenon accesible.

 Other possibility for observing the phenomenon could be to reduce the effective interacion using Feschbasch 
 resonances \cite{Cornell}  until a level at which the vortex would become ``unstable". This procedure may be 
 experimentally controlled with high precision and very low loss of atoms (e.g using Rb$^{87}$ the JILA group has been
  able to reach a regime of zero effective interactions \cite{Feschbach2}). In the later case the preparation of a vortex 
  should not be a problem since it could be generated by using either a rotating trap or a multiple condensate system  
  and subsequent evaporation of one of the species  \cite{Cornell2}.

The lack of structural stability for vortices is a surprising result. It has
been widely accepted that the existence of a macroscopic wave-function is a
sufficient condition for the existence and stability of vortices no matter what
the actual model is. The dynamical stability of vortices as seen in experiments
and in numerical studies with symmetric \cite{Fetter-review,vortices} and
asymmetric \cite{feder} vortices in strongly interacting condensates, and some
results with symmetric vortices in weakly interacting condensates \cite{butts},
has lead us to consider vortices and their topological charge as robust
entities.

Nevertheless, as we have shown in this work there exists a mechanism which changes the topological
charge of the vortex while preserving the energy of the cloud. This mechanism
is completely different from the dissipative mechanisms usually considered in
BEC \cite{rokhsar}, which imply loss of energy and induce a spiraling of the
vortex out of the condensate.  In fact, the mechanism presented here is
reminiscent of a phenomenon recently discovered in light beams \cite{torner},
where vortices where observed to undergo extremely sharp Berry trajectories
that, in the experimental plane of observation, appear as an edge-line
dislocation which eventually yields the inversion of the vortex charge.

These predictions, striking as they are, shed light on the essential role played by
nonlinearity in these systems and it is another manifestation of the richness
of the nonlinear quantum phenomena appearing in BEC.

This work has been supported by grants BFM2000-0521 and TIC2000-1010.


\begin{thebibliography}{99}
\bibitem{note}{In other contexts the term ``weakly interacting condensates" is used to refer to
    gaseous condensates as oposed to ``strongly interacting condensates" or
    liquid condensates made of He-II. Notice that we always refer to gaseous
    condensates and our use of weak and strong corresponds only to
    the value of the relevant parameter $UN$.}
\bibitem{note2}{This effectively two--dimensional system should not be confused with a 
purely two--dimensional system, where the physics will probably be very different. See e.g.
the discussion in Ref. \onlinecite{MIT-new}.}
\end{thebibliography}
\end{document}